\documentstyle[12pt,epsfig]{article}
\newcommand{\Z}{{\sf Z \!\!\! Z}}
\newcommand{\N}{{\sf I \!\! N}}
\newcommand{\R}{{\sf I \!\! R}}

\makeatletter
\@addtoreset{equation}{section}
\makeatother

\title{Lattice Fluid Dynamics from Perfect Discretizations of Continuum Flows
\footnote{This work is supported in part by funds provided by the U.S.
Department of Energy (D.O.E.) under cooperative research agreement
DE-FC02-94ER40818.}}
\author{E. Katz and U.-J. Wiese \\ \\
Center for Theoretical Physics, \\
Laboratory for Nuclear Science, and Department of Physics \\
Massachusetts Institute of Technology (MIT) \\
Cambridge, Massachusetts 02139, U.S.A. \\ \\
MIT Preprint, CTP 2423 \\ \\}
 
\begin{document}
\maketitle
\begin{abstract} \normalsize
 
We use renormalization group methods to derive equations of motion for large
scale variables in fluid dynamics. The large scale variables are averages
of the underlying continuum variables over cubic volumes, and naturally live 
on a lattice. The resulting lattice dynamics represents a perfect 
discretization of continuum physics, i.e. grid artifacts are completely
eliminated. Perfect equations of motion are derived 
for static, slow flows of incompressible, viscous fluids.
For Hagen-Poiseuille flow in a channel with square cross section the 
equations reduce to a perfect discretization of the Poisson equation 
for the velocity field with Dirichlet boundary conditions. The perfect large 
scale Poisson equation is used in a numerical simulation, and is shown to
represent the continuum flow exactly. For non-square cross sections we use a
numerical iterative procedure to derive flow equations that are approximately
perfect.

\end{abstract}
 
\maketitle
 
\newpage

\section{Introduction}

Problems of fluid flow are of great practical importance in such areas as
mechanical engineering and meteorology. A full theoretical understanding
of fluid flow is very difficult because of the complex dynamics. In particular,
in the regime of fully developed turbulence we face a dynamical system with
fluctuations on all length scales. The small scale structures influence the
dynamics at larger scales in a way that is hard to control analytically. 
Numerical simulations are therefore an essential tool in fluid dynamics. 
However, with present day computers a full simulation of realistic problems 
from first principles is beyond reach.
Already simple model problems exhaust the biggest machines available. Hence,
in practical applications one makes additional assumptions, especially about 
the small scale dynamics, which are difficult to justify theoretically, and
which involve free parameters. In fact, the primary systematic error in
numerical simulations of fluid flow is due to the finite grid size that is
necessary to discretize the continuum Navier-Stokes equations. 

The main
purpose of this paper is to point out that perfect discretizations, i.e. ones
that are completely free of grid artifacts, exist and can be constructed 
explicitly. The existence of perfect discretizations is a direct consequence
of Wilson's renormalization group theory \cite{Wil74}.
The idea of the renormalization
group is to deal with the fluctuations at different length scales step by
step in scale. In this way the influence of small scale fluctuations on the
large scale dynamics can, at least in principle, be controlled exactly.
The renormalization group approach has been useful in the study
turbulence and has been applied to various problems previously (see
\cite{Ors1}, \cite{Ors2}, \cite{star} for some examples).

Recently, however, in the context of lattice quantum field theories so-called perfect actions
have been constructed for various models using the renormalization group. Originally, Hasenfratz and 
Niedermayer iterated a renormalization group transformation to obtain a
perfect discretization for the 2-dimensional $O(3)$ nonlinear $\sigma$-model, 
and they demonstrated that grid artifacts were eliminated even on coarse lattices 
\cite{Has94}. The same method has been applied to lattice fermions
\cite{Wie93,Bie95}, to 4-dimensional pure $SU(3)$ gauge theory \cite{DeG95} 
and to full QCD \cite{Bie95a}. Here we use similar methods in classical physics.
In principle, the renormalization group can be used to derive perfect 
discretizations for any differential equation. Examples of great practical 
importance are the Laplace and Poisson equations, Maxwell's equations and the 
Navier-Stokes equation. A perfect discretization of the Laplace equation was
obtained by Bell and Wilson in an early study of scalar lattice field theory
\cite{Bel75}. The perfect action for a classical gauge field --- and hence a 
perfect discretization of Maxwell's equations --- has been constructed in 
refs.\cite{DeG95,Bie95a}. In this work we concentrate on the Navier-Stokes
equations. Here we do not treat the problem in its full complexity yet. 
Specifically, we restrict ourselves to incompressible fluids and we assume 
static, slow flows.

The first step in our program is to define large scale lattice variables from
the underlying continuum fields. In the case of the Navier-Stokes equations
these are the velocity and pressure fields. We average the continuum pressure
over a cube of side length $a$ to define the large scale pressure variables,
which then naturally live on a lattice of spacing $a$ consisting of the
cube centers. Similarly, the velocity field is averaged over the interface
between two neighboring cubes, and hence the corresponding lattice field
lives on the links connecting neighboring lattice points. This construction
ensures that the continuity equation maintains a simple form on the lattice.
By construction we know exactly how the
large scale variables are related to the continuum fields. The goal of the
next step is to derive equations of motion for the large scale fields.
Solving these equations yields exact results for the averaged continuum
quantities. This is in contrast to standard discretization procedures, where
the lattice field represents an approximation of the continuum field at the
same point. The approximation becomes exact only in the continuum limit
$a \rightarrow 0$, because the value of the field at a specific point is
influenced by the dynamics at arbitrarily small scales. A perfect 
discretization, on the other hand, ensures exact results for averaged 
continuum quantities already at finite $a$. Still, the values of the continuum 
fields can be reconstructed from the lattice data by using so-called perfect 
fields.

The second step is to derive exact equations of motion for the lattice fields.
One possibility to do that is to iterate a discrete renormalization group
transformation. This is done by starting on a very fine lattice (in fact,
at the end the lattice spacing of the fine lattice is sent to zero), and then
performing an infinite sequence of blocking steps, in which the lattice fields
are averaged over blocks, 
and effective equations of motion for the block variables are
derived. Here we proceed differently by blocking directly from the continuum.
The result is identical to the discrete, iterative procedure, but the 
derivation is more transparent. Wilson was the first to discover this
technique \cite{Wil76}, which has not been very well known until recently.
For example, blocking from the continuum was also used in the derivation of 
the QCD perfect action \cite{Bie95a}. In contrast to quantum field theory
there is no action for the full Navier-Stokes problem. Still, blocking from
the continuum can be directly applied to the equations of motion. The 
constraint that identifies the large scale lattice variables with the averaged
continuum fields is implemented through a Lagrange multiplier. Solving the 
equations
for the continuum variables in terms of the lattice fields and plugging the
result back into the constraint equations leads to the equations for the large
scale variables. 

For numerical applications it is essential that the lattice equations of
motion are as local as possible. In this paper we distinguish three types
of locality: ultralocal, local and nonlocal. A lattice equation is called 
ultralocal if it couples the lattice field at a given site to a finite number 
of neighboring sites only. Consider, for example, the 1-dimensional Laplace 
equation 
\begin{equation}
\frac{d^2 f(x)}{d x^2} = 0.
\end{equation}
Its standard lattice discretization
\begin{equation}
\label{standard}
\frac{1}{a^2}[f(x+a) - 2 f(x) + f(x-a)] = 0
\end{equation}
is ultralocal because it couples the field at the point $x$
to its nearest neighbors at $x+a$ and $x-a$ only. In general, a perfect
discretization will not lead to an ultralocal coupling. The couplings
to fields at distant sites are non-zero, but exponentially suppressed. In that
case we say that the lattice equation is local. By optimizing the parameters 
of the renormalization group transformation one can control the strength of
the exponential decay. It turns out that the parameters can be chosen such 
that the perfect equations become ultralocal in one dimension. In practice, 
even in higher dimensions this leads to extremely local perfect lattice 
equations, whose exponentially suppressed couplings to distant 
neighbors can safely be neglected. In particular, as it will become clear
later, even the ultralocal eq.(\ref{standard}) turns out to be perfect.
This would not be the case for the common approach to discretization problems,
where one interprets lattice derivatives as approximations of continuum 
derivatives at the same point.
Still, even then one can find a perfect discretization. It is a well
known result in engineering that a smooth function $f(x)$ can be reconstructed
from its values $f(n a)$ on a lattice by
\begin{equation}
f(x) = \sum_{n \in \Z} f(n a) \frac{\sin \pi(x/a - n)}
{\pi (x/a - n)}.
\end{equation}
Here we have assumed that $f(x)$ is infinitely differentiable and that its 
Fourier transform exists. The above equation yields
\begin{equation}
\frac{d^2 f(m a)}{dx^2} = - \frac{1}{a^2} \Big[\frac{\pi^2}{3} f(0) + 
2 \sum_{n \in \Z \backslash \{ m \} } f(n a) \frac{(-1)^{n-m}}{(n-m)^2}\Big].
\end{equation}
This exactly represents the continuum second derivative of the function at
the lattice point $m a$. However, now the lattice derivative comprises terms
at arbitrarily large distances, and their contributions are suppressed only
power-like. This is what we call a nonlocal coupling. In this case the
contributions from large distances cannot be neglected and therefore this
equation is not of practical use. In contrast, the
advantage of using block averaged fields is that their perfect discretizations
are local.

As opposed to previous applications of perfect discretizations, which dealt
with infinite volumes or finite systems with periodic boundary conditions, 
the incorporation of more general boundary conditions is essential
in fluid dynamics. While it is straightforward to implement periodic
boundary conditions, it is more difficult to treat fixed boundary conditions,
which occur in typical fluid dynamics applications. Still, it turns out that
fixed boundary conditions can be represented exactly in the perfect lattice
equations. Here we restrict ourselves to channels whose cross section is a
rectangular polygon.
Arbitrarily curved boundaries can in principle be handled using curvelinear
coordinates, but we do not elaborate on this issue here. 

The paper is organized as follows. In section 2 we introduce the large scale
lattice pressure and velocity variables by blocking from the continuum, and
we investigate their behavior under discrete renormalization group
transformations. Section 3 contains the derivation of perfect equations of
motion for Hagen-Poiseuille flow. The parameter in the renormalization group
transformation is optimized for ultralocality in two dimensions. We verify
explicitly that also in three dimensions continuum physics is exactly
reproduced. In section 4 we show analytically that the equations of
motion indeed satisfy a factor 2 renormalization group step.  Section 5
presents results of a numerical calculation of the lattice fields.  In
section 6 we construct the perfect velocity field that allows
reconstruction of the continuum field from the lattice data. Section 7 deals
with static, slow flows in general. We derive the perfect lattice version
of the corresponding Navier-Stokes equations.  In section 8 we present a
perturbative method of introducing the non-linear term to equation of motion.
Finally, section 9 contains concluding remarks on how to extend our 
construction to the full Navier-Stokes equations and to arbitrarily shaped
boundaries. We also speculate about applications of perfect discretizations
to numerical simulations of turbulent flows. 

\section{Large scale lattice variables}

\begin{figure}[hbt]
\label{blocking}
\centering
\epsfig{file=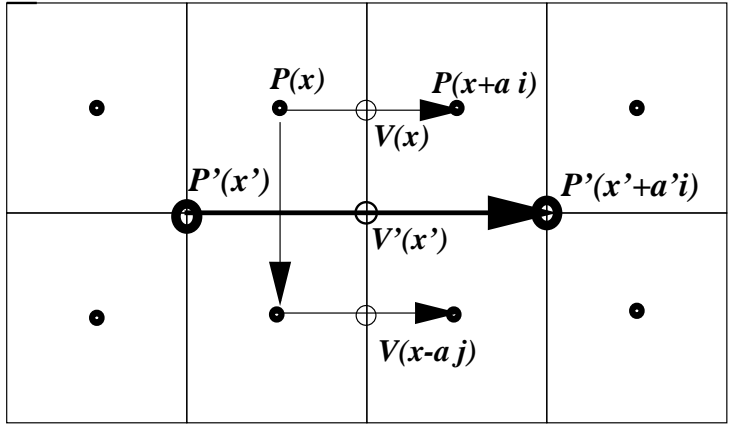}
\caption{\it The geometry of the block factor 2 renormalization group
transformation.}
\end{figure}

The continuum Navier-Stokes equations are formulated in terms of pressure
and velocity variables. Here we consider static flows in $d$ dimensions, and 
hence pressure $p(y)$ and velocity $v_i(y)$ with $i \in \{1,2,...,d \}$ are 
functions of the position vector $y$ only. For incompressible fluids the 
velocity field obeys the continuity equation
\begin{equation}
\partial_i v_i(y) = 0.
\end{equation}
To define the large scale lattice pressure variables $P_x$ the continuum
pressure field $p(y)$ is averaged over a $d$-dimensional cube $c_x$ 
of side length $a$ centered at $x$,
\begin{equation}
P_x = \frac{1}{a^d} \int_{c_x} d^dy \ p(y).
\end{equation}
The variables $P_x$ naturally live on a lattice of spacing $a$ formed by
the cube centers $x$. Similarly, for the velocity field we define
\begin{equation}
V_{i,x} = \frac{1}{a^{d-1}} \int_{f_{i,x}} d^{d-1}y \ v_i(y),
\end{equation}
where $f_{i,x}$ is the $(d-1)$-dimensional face separating the cubes
$c_{x-a\hat i/2}$ and $c_{x+a\hat i/2}$. Note that $\hat i$ is the unit
vector in the $i$-direction. It is natural to associate the velocity
variables with the links connecting neighboring lattice sites $x-a\hat i/2$
and $x+a\hat i/2$. Now the $i$-component of $x$ is a half-integer multiple
of the lattice spacing. The lattice velocity variables can be interpreted
as volume flow rates per cross sectional area of the faces separating
neighboring cubes. This definition
of the lattice velocity field is natural because the lattice continuity
equation then assumes an ultralocal form. It is simply given by
\begin{eqnarray}
\delta V_x&=&\sum_i [V_{i,x+a\hat i/2} - V_{x-a\hat i/2}]=
\frac{1}{a^{d-1}} \sum_i \Big[ \int_{f_{i,x+a\hat i/2}} \!\!\!\!\!\!\!\!\!\!
d^{d-1}y \ v_i(y)
- \int_{f_{i,x-a\hat i/2}} \!\!\!\!\!\!\!\!\!\!
d^{d-1}y \ v_i(y) \Big] \nonumber \\
&=&\frac{1}{a^{d-1}} \int_{\partial c_x} d^{d-1}\sigma_i \ v_i(y) =
\frac{1}{a^{d-1}} \int_{c_x} d^dy \ \partial_i v_i(y) = 0.
\end{eqnarray}
Here we have used Gauss' law together with the continuum continuity equation.
The definition of lattice variables by blocking from the continuum can be
viewed as a renormalization group transformation with an infinite blocking 
factor. In fact, the same result could be obtained by iterating block factor
2 transformations starting from an arbitrarily fine lattice. To identify the
corresponding factor 2 transformation we now block from the continuum to a
lattice of twice the lattice spacing $a' = 2 a$. The continuum pressure
variables are then integrated over cubes $c'_{x'}$ of side length $a'$ which
contain $2^d$ original cubes $c_x$. Hence, the resulting pressure variable
\begin{equation}
P'_{x'}=\frac{1}{{a'}^d} \int_{c'_{x'}} d^dy \ p(y) = 
\frac{1}{2^d} \sum_{x \in x'} \frac{1}{a^d} \int_{c_x} d^dy \ p(y) =
\frac{1}{2^d} \sum_{x \in x'} P_x
\end{equation}
on the coarse lattice is a block average of the pressure variables on the
fine lattice. We use $x \in x'$ to denote that the cube $c_x$ belongs to the
block $c'_{x'}$. The corresponding geometry is illustrated in figure 
\ref{blocking}.
Similarly, for the velocity field we have
\begin{eqnarray}
V'_{i,x'}&=&\frac{1}{{a'}^{d-1}} \int_{f'_{i,x'}} d^{d-1}y \ v_i(y) 
\nonumber \\
&=&\frac{1}{2^{d-1}} \sum_{x \in x'} \frac{1}{a^{d-1}} \int_{f_{i,x}} d^{d-1}y
\ v_i(y) = \frac{1}{2^{d-1}} \sum_{x \in x'} V_{i,x}.
\end{eqnarray}
Here $x \in x'$ denotes that the sum includes faces $f_{i,x}$ on the fine
lattice that belong to the face $f'_{i,x'}$ on the coarse lattice. Of course,
by construction the coarse lattice velocity variables also obey the 
continuity equation. This block factor 2 renormalization group
transformation (RGT)
will be used later to demonstrate that the structure of the perfect equations 
of motion reproduces itself under renormalization.

\section{Perfect equations of motion for Hagen-Poiseuille flow}

For simplicity we first discuss the derivation of perfect equations of motion
in the context of Hagen-Poiseuille flow. The general equations for static,
slow flows are derived in section 6. We start from the Navier-Stokes equation
\begin{equation}
\partial_t v_i(y) + (v_j(y) \partial_j) v_i(y) = 
- \frac{1}{\rho} \partial_i p(y) + \nu \partial_j \partial_j v_i(y).
\end{equation}
Here $\nu$ is the kinematic viscosity of the fluid, and $\rho$ is its
density. We restrict ourselves to incompressible flows, i.e. to a constant
density, and thus the continuity equation applies. The appropriate boundary
condition for a viscous fluid at a fixed wall is $v_i(y) = 0$ for all $i$.
Furthermore, we consider
static Hagen-Poiseuille flow, i.e. a constant pressure gradient $\delta p$,
in a channel with square cross section $A=L^{d-1}$ along the $d$-direction. 
Hence,
the pressure is given by $p(y) = \delta p \ y_d$. The velocity is along the
channel, i.e. only $v_d(y)$ is nonzero and in addition independent of $y_d$.
Then the equations reduce to the $(d-1)$-dimensional Poisson equation
\begin{equation}
\partial_i \partial_i v_d(y) = \frac{\delta p}{\nu \rho},
\end{equation}
with Dirichlet boundary conditions $v_d(y) = 0$ at the edge of the square.
This equation follows from a variational principle of an action
\begin{equation}
S[v_d] = \int_A d^{d-1}y [\frac{1}{2} \partial_i v_d(y) \partial_i v_d(y) + 
\frac{\delta p}{\nu \rho} v_d(y)].
\end{equation}
Note that here $i$ runs from 1 to $(d-1)$ only.  The action is useful in our
derivation of the perfect equations of motion. Yet we want to stress that
the existence of an action is not really necessary. This is crucial because
there is no action principle for the full Navier-Stokes problem.  In
fact, writing the above action was possible only because we have removed
the time dependent dissipative element of the full problem and have
forced the flow to be static.

What is the perfect discretization of the Poisson equation? To answer that
question we consider the large scale lattice variable
\begin{equation}
\label{constraint}
V_{d,x} = \frac{1}{a^{d-1}} \int_{f_{d,x}} d^{d-1}y \ v_d(y).
\end{equation}
Note that the face $f_{d,x}$ is nothing more than a $(d-1)$-dimensional cube,
and $V_{d,x}$ behaves practically as a $(d-1)$-dimensional scalar. We want
to impose eq.(\ref{constraint}) as a constraint for each face $f_{d,x}$. 
For this purpose we add it to the action using a Lagrange multiplier 
field $\eta_{d,x}$ that lives on the face $f_{d,x}$
\begin{eqnarray}
S[v_d,V_d,\eta_d]&=&\int_A d^{d-1}y 
[\frac{1}{2} \partial_i v_d(y) \partial_i v_d(y) +
\frac{\delta p}{\nu \rho} v_d(y)] \nonumber \\
&+&\sum_x a^{d-1} [\eta_{d,x} 
(V_{d,x} - \frac{1}{a^{d-1}} \int_{f_{d,x}} d^{d-1}y \ v_d(y))
- \frac{a^2\alpha}{2} \eta_{d,x}^2].
\end{eqnarray}
We have also included a term quadratic in the Lagrange multiplier field.
This would not really be necessary, but its coefficient $\alpha$ will allow
us to optimize the perfect equation's locality, and therefore it is of great
practical importance. Next we derive equations of motion from the variational
principle of the action. Varying the action with respect to $v_d(y)$ yields
\begin{equation}
\label{Navierv}
\partial_i \partial_i v_d(y) - \frac{\delta p}{\nu \rho} + 
\sum_x \eta_{d,x} \theta_{d,x}(y) = 0,
\end{equation}
where $\theta_{d,x}(y) = 1$ for $y \in f_{d,x}$ and zero otherwise. Varying the
action with respect to $\eta_{d,x}$ one obtains
\begin{equation}
\label{Navier}
V_{d,x} - \frac{1}{a^{d-1}} \int_{f_{d,x}} d^{d-1}y \ v_d(y) 
- a^2 \alpha \eta_{d,x} = 0.
\end{equation}
It appears that the constraint of eq.(\ref{constraint}) 
is correctly implemented
only for $\alpha = 0$. However, varying the action with respect to $V_{d,x}$
gives
\begin{equation}
\eta_{d,x} = 0.
\end{equation}
To solve the above equations we now go to momentum space. Since the velocity
vanishes at the boundary the appropriate Fourier transform takes the form
\begin{equation}
v_d(k) = \int_A d^{d-1}y \ v_d(y) \prod_{i=1}^{d-1} \sin(k_i y_i).
\end{equation}
Here the
\begin{equation}
\label{momentum}
k_i = \frac{\pi}{L} m_i,
\end{equation}
are discrete momenta with $m_i \in \N$. Note that $v_d(k)$ extends naturally
to negative momentum values because
\begin{equation}
v_d(k_1,...,-k_i,...,k_{d-1}) = - v_d(k_1,...,k_i,...,k_{d-1}).
\end{equation}
Similarly, for the lattice velocity field we have
\begin{equation}
V_d(k) = a^{d-1} \sum_x V_{d,x} \prod_{i=1}^{d-1} \sin(k_i y_i). 
\end{equation} 
Now the sum is finite and extends over $m_i = 1,2,...,N$ only, where $N = L/a$.
The Lagrange multiplier field $\eta_{d,x}$ is transformed analogously. 
In Fourier space the action takes the form
\begin{eqnarray}
&&S[v_d,V_d,\eta_d]=\frac{1}{a^{d-1}}\sum_{k \in (\frac{\pi N}{L})^{d-1}} [\frac{1}{2} k^2 v_d(k)^2
+ \frac{\delta p}{\nu \rho} \delta_L(k) v_d(k)] \nonumber \\
&&+\frac{1}{a^{d-1}}\sum_k [(V_d(k) - \sum_{l \in \Z^{d-1}} v_d(k + 2 \pi l/a)
\Pi_d(k + 2 \pi l/a) \prod_{i=1}^{d-1} (-1)^{l_i}) \eta_d(k) \nonumber \\-
&&\frac{a^2 \alpha}{2} \eta_d(k)^2]. \nonumber \\ \
\end{eqnarray}
The second sum over $k$ is restricted to $m_i \in \{1,2,...,N\}$. In the
above expression we have introduced
\begin{equation}
\delta_L(k) = \prod_{i=1}^{d-1} \frac{1 - (-1)^{m_i}}{k_i}
\end{equation}
as the Fourier transform of the constant 1. The function
\begin{equation}
\Pi_d(k) = \prod_{i=1}^{d-1} \frac{2 \sin(k_i a/2)}{k_i a}
\end{equation}
results from the Fourier transform of $\theta_{d,x}(y)$.  
In Fourier space eq.(\ref{Navierv}) takes the form
\begin{equation}
\label{equationofmotion}
- k^2 \ v_d(k) - \frac{\delta p}{\nu \rho} \delta_L(k) + \Pi_d(k) \eta_d(k) = 0.
\end{equation}
In eq.(\ref{equationofmotion}) the momentum $k$ has the form of 
eq.(\ref{momentum}) with $m_i \in N$ being an arbitrary integer. On the
other hand, as it stands $\eta_d(k)$ is defined only for momenta with
$m_i \leq N$. For other momenta $\eta_d(k)$ naturally extends to
\begin{eqnarray}
&&\eta_d(k_1,...,-k_i,...,k_{d-1}) = - \eta_d(k_1,...,k_i,...,k_{d-1}), 
\nonumber \\
&&\eta_d(k + 2 \pi l_i \hat i/a)=(-1)^{l_i} \eta_d(k).
\end{eqnarray}
Solving eq.(\ref{equationofmotion}) yields
\begin{equation}
\label{vequation}
v_d(k) = \frac{1}{k^2}[\Pi_d(k) \eta_d(k) - \frac{\delta p}{\nu \rho} 
\delta_L(k)].
\end{equation}
We plug this back into the action and obtain
\begin{equation}
S[V_d,\eta_d]=\frac{1}{a^{d-1}}\sum_k [- \frac{1}{2} \eta_d(k) \omega_{dd}(k)^{-1} \eta_d(k)
+ \frac{\delta p}{\nu \rho} \Delta_L(k) \eta_d(k) + 
V_d(k) \eta_d(k)],
\end{equation}
where
\begin{eqnarray}
\label{omega}
\omega_{dd}(k)^{-1}&=&\sum_{l \in \Z^{d-1}} \frac{1}{(k + 2 \pi l/a)^2}
\Pi(k + 2 \pi l/a)^2 + \alpha a^2 \nonumber \\
\Delta_L(k)&=&\sum_{l \in \Z^{d-1}} \delta_L(k + 2 \pi l/a) 
\frac{1}{(k + 2 \pi l/a)^2} \Pi_d(k + 2 \pi l/a) (-1)^{\sum_i l_i}.
\end{eqnarray}
Minimizing with respect to the auxiliary field $\eta_d(k)$ yields
\begin{equation}
\label{etaequation}
\eta_d(k) = \omega_{dd}(k)(V_d(k) 
+ \frac{\delta p}{\nu \rho} \Delta_L(k)).
\end{equation}
Reinserting this into the action finally gives
\begin{equation}
S[V_d] = \frac{1}{a^{d-1}}\sum_k [\frac{1}{2} V_d(k) \omega_{dd}(k) V_d(k) +
 \omega_{dd}(k) \frac{\delta p}{\nu \rho} \Delta_L(k) V_d(k)].
\end{equation}
This is a perfect action for the Poisson equation in a finite volume. The
resulting perfect equation of motion takes the form
\begin{equation}
\label{perfecteq}
\omega_{dd}(k) V_d(k) +
\omega_{dd}(k) \frac{\delta p}{\nu \rho} \Delta_L(k) = 0.
\end{equation}
By construction it is clear that this equation perfectly represents continuum
physics. It was helpful to have an action whose variation produced the perfect
equation of motion, but the action was not really necessary. As we will see
later, all we really need are the equations of motion. This is important for
the full Navier-Stokes problem, for which an action does not exist.

For practical applications eq.(\ref{perfecteq}) needs to be transformed back
to coordinate space. It should be noted that $\omega_{dd}(k)$ contains the
arbitrary parameter $\alpha$. Hence, eq.(\ref{perfecteq}) represents a whole
family of perfect equations of motion. Now we want to optimize $\alpha$ such
that the equations become as local as possible. For this purpose we consider
$d=2$. Then the sums in eq.(\ref{omega}) can be performed analytically, and
one obtains
\begin{eqnarray}
\omega_{22}(k_1)^{-1}&=&a^2[\frac{1}{4 \sin^2(a k_1/2)} - \frac{1}{6}
+ \alpha]
\nonumber \\
\Delta_L(k_1)&=& a^3[\frac{1}{8 \sin^3(a k_1/2)} - \frac{1}{6} 
\frac{1}{2 \sin(a k_1/2)}](1 - (-1)^{m_1}),
\end{eqnarray}
where again $k_1=\pi m_1/L$.

For the choice of $\alpha = 1/6$ this gives 
$\omega_{22}(k_1) = 4/a^2 \sin^2(a k_1/2)$, which corresponds to the standard
ultralocal second derivative in coordinate space. Transforming the whole perfect
equation of motion eq.(\ref{perfecteq}) to coordinate space one obtains
\begin{equation}
\label{perfect2d}
\frac{1}{a^2} [V_{2,x_1+a} - 2 V_{2,x_1} + V_{2,x_1-a}] = 
\frac{\delta p}{\nu \rho},  
\end{equation}
for $x_1 = (n - 1/2)a$ with $n \neq 1, N$, i.e. for points away from the
boundary. At the left boundary one finds
\begin{equation}
\frac{1}{a^2} [V_{2,3a/2} - 3 V_{2,a/2}] = 
\frac{2}{3} \frac{\delta p}{\nu \rho},
\end{equation}
and there is an analogous expression at the right boundary. It is straightforward to verify that the averaged continuum solution does in fact satisfy
these equations. In the continuum the Hagen-Poiseuille flow in $d=2$ is
given by
\begin{equation}
v_2(y_1) = \frac{1}{2} \frac{\delta p}{\nu \rho} y_1(y_1 - L).
\end{equation}
Consequently, the corresponding large scale lattice variable takes the form
\begin{equation}
V_{2,x_1} = \frac{1}{a} \int_{x_1-a/2}^{x_1+a/2} dy_1 \ v_2(y_1)
= \frac{1}{2} \frac{\delta p}{\nu \rho} [x_1(x_1 - L) + \frac{a^2}{12}],
\end{equation}
which indeed solves eq.(\ref{perfect2d}). 

In $d=3$ the perfect equations of motion are no longer ultralocal. Still,
when we use the optimized $\alpha = 1/6$ from $d=2$ also in 3 dimensions
the equations turn out to be extremely local. In coordinate space the
equation of motion takes the form
\begin{equation}
\label{coordseqn}
\sum_{x'} \omega_{dd}(x,x') V_{d,x'} = 
- \frac{\delta p}{\nu \rho} \Delta_L(x),
\end{equation}
where $\omega_{dd}(x,x')$ and $\Delta_L(x)$ are Fourier transforms of the
corresponding quantities in momentum space. Their values for a 3-d system with
$N=5$ , $a=1$ are given in tables \ref{table1}, \ref{table2} and \ref{table3} for various values of $x$ and 
plots for the same values are shown in figures 2, \ref{fig2}, and
4. Also the values for $D_L(x)$ (with $\delta p/\nu \rho$ 
set to 1) are provided in table
\ref{Tdel}, and displayed in Fig. 5.

\begin{table}
\centering
\begin{tabular}{|c|c|c|c|c|c|}
\hline
$V_{(1/2,1/2)}$ & $x=1/2$ & $x=3/2$ & $x=5/2$ & $x=7/2$ & $x=9/2$ \\
\hline
$y=1/2$ & 4.28599 & -0.42637 & -0.00118 & -0.00013 & -0.00000 \\
\hline
$y=3/2$ & -0.42637 & -0.18734 & -0.00227 & -0.00003 & -0.00000 \\
\hline
$y=5/2$ & -0.00118 & -0.00227 & 0.00151 & 0.00006 & 0.00000 \\
\hline 
$y=7/2$ & -0.00013 & -0.00003 & 0.00006 & -0.00001 & 0.00000 \\
\hline
$y=9/2$ & -0.00000 & -0.00000 & 0.00000 & 0.00000 & 0.00000 \\
\hline
\end{tabular}
\caption{\it Coupling of a corner field}
\label{table1}
\end{table}

\begin{figure}[hbt]
\label{fig1}
\centering
\epsfig{file=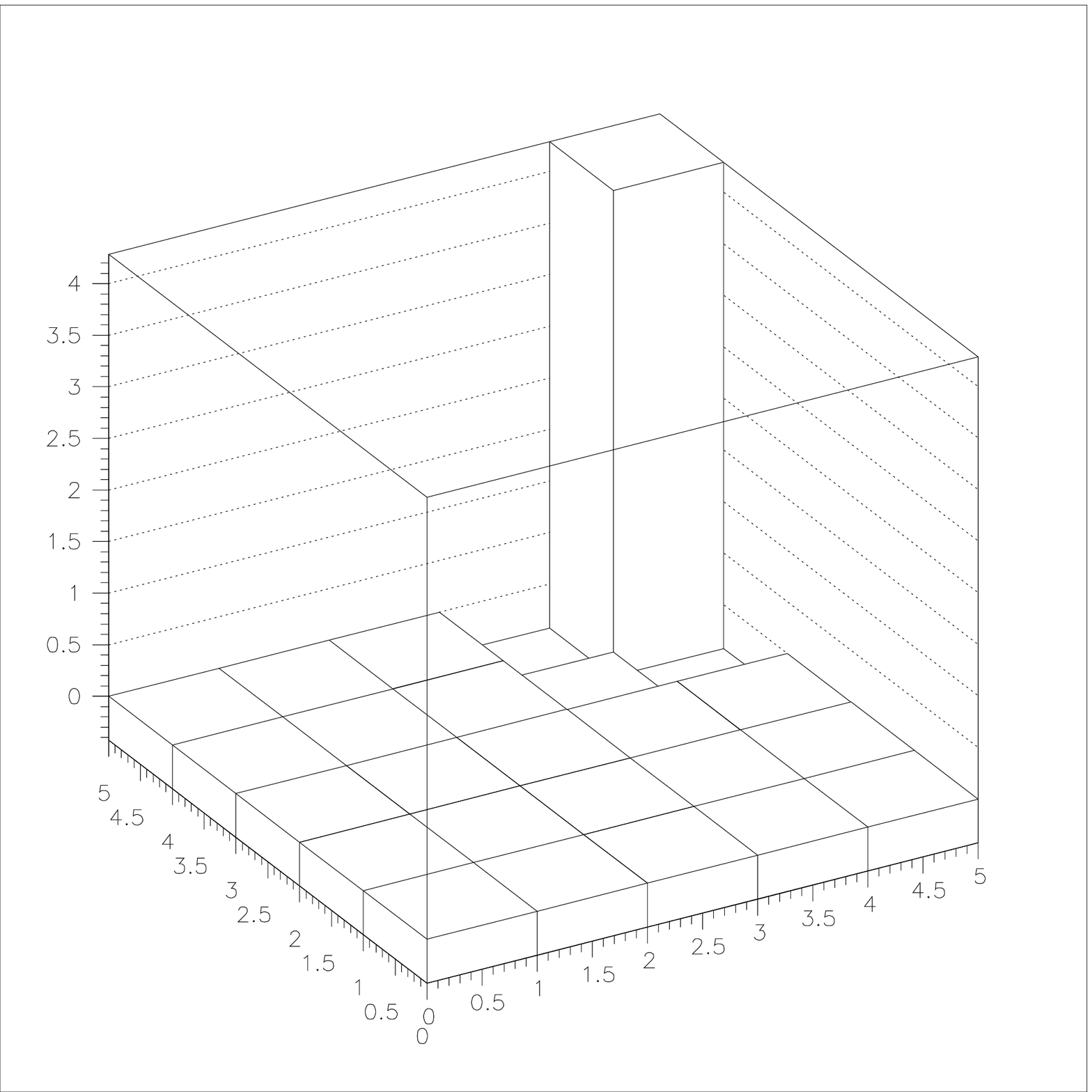, width=420pt}
\caption{\it The coupling of a corner field to the lattice}
\end{figure}

\begin{table}
\centering
\begin{tabular}{|c|c|c|c|c|c|}
\hline
$V_{(5/2,5/2)}$ & $x=1/2$ & $x=3/2$ & $x=5/2$ & $x=7/2$ & $x=9/2$ \\
\hline
$y=1/2$ & 0.00151 & -0.00070 & -0.00188 & -0.00070 & 0.00151 \\
\hline
$y=3/2$ & -00070 & -0.19033 & -0.61801 & -0.19033 & -0.00070 \\
\hline
$y=5/2$ & -0.00188 & -0.61801 & 3.24027 & -0.61801 & -0.00188 \\
\hline 
$y=7/2$ & -0.00070 & -0.19033 & -0.61801 & -0.19033 & -0.00070 \\
\hline
$y=9/2$ & 0.00151 & -0.00070 & -0.00188 & -0.00070 & 0.00151 \\
\hline
\end{tabular}
\caption{\it Coupling of a center field}
\label{table2}
\end{table}

\begin{figure}[hbt]
\label{fig2}
\centering
\epsfig{file=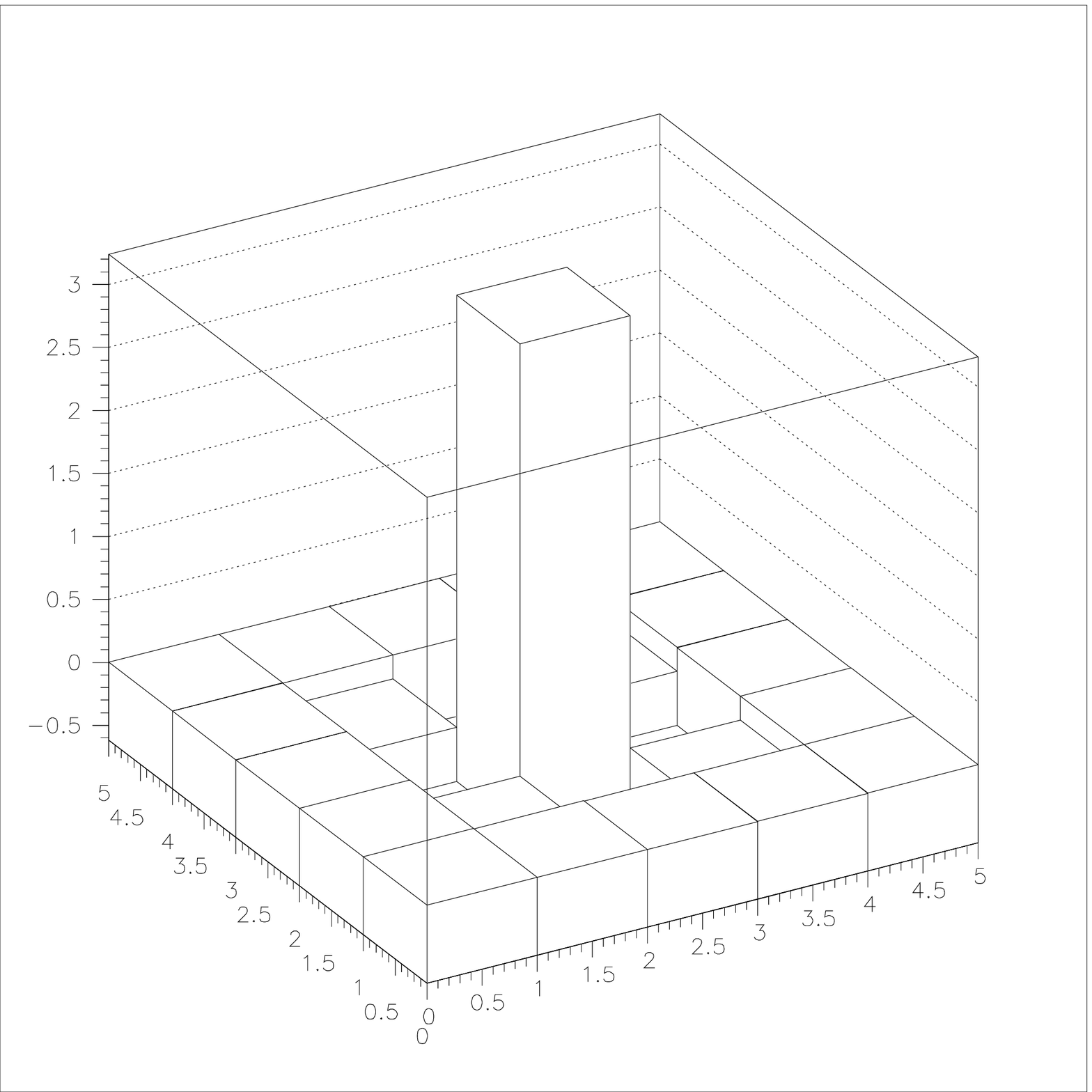,width=420pt}
\caption{\it The coupling of a center field to the lattice}
\end{figure}

\begin{table}
\centering
\begin{tabular}{|c|c|c|c|c|c|}
\hline
$V_{(5/2,1/2)}$ & $x=1/2$ & $x=3/2$ & $x=5/2$ & $x=7/2$ & $x=9/2$ \\
\hline
$y=1/2$ & -0.00118 & -0.42769 & 3.85830 & -0.42769 & -0.00118 \\
\hline
$y=3/2$ & -0.00227 & -0.18965 & -0.61602 & -0.18965 & -0.00227 \\
\hline
$y=5/2$ & -0.00151 & -0.00070 & -0.00188 & -0.00070 & -0.00151 \\
\hline 
$y=7/2$ & 0.00006 & 0.00002 & -0.00011 & 0.00002 & 0.00006 \\
\hline
$y=9/2$ & 0.00000 & 0.00000 & -0.00000 & 0.00000 & 0.00000 \\
\hline
\end{tabular}
\caption{\it Coupling of an edge field}
\label{table3}
\end{table}

\begin{figure}[hbt]
\label{fig3}
\centering
\epsfig{file=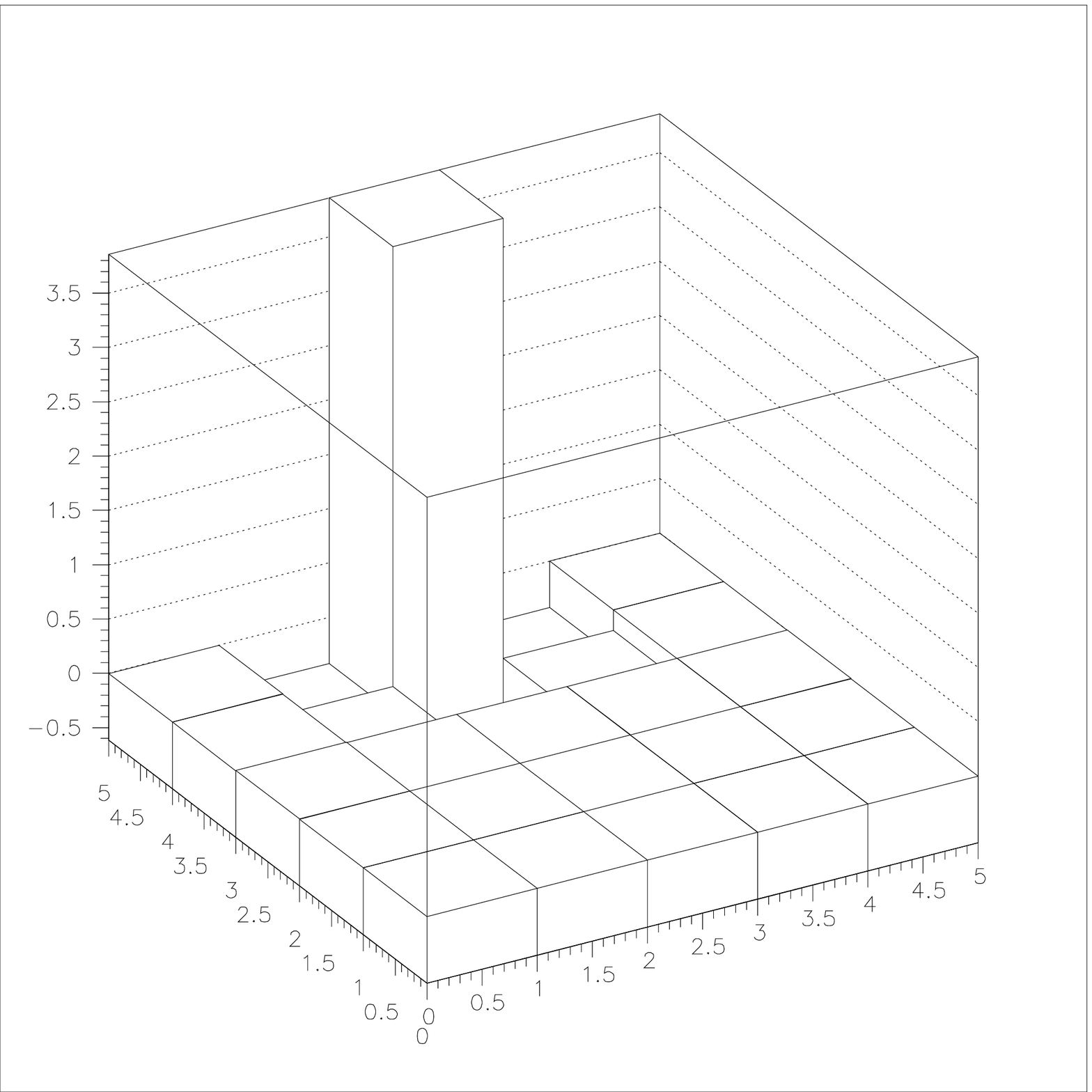,width=420pt}
\caption{\it The coupling of an edge field to the lattice}
\end{figure}

\begin{table}
\centering
\begin{tabular}{|c|c|c|c|c|c|}
\hline
* & $x=1/2$ & $x=3/2$ & $x=5/2$ & $x=7/2$ & $x=9/2$ \\
\hline
$y=1/2$ & 0.45985 & 0.66567 & 0.66624 & 0.66567 & 0.45985 \\
\hline
$y=3/2$ & 0.66567 & 0.99806 & 0.99902 & 0.99806 & 0.66567 \\
\hline
$y=5/2$ & 0.66624 & 0.99902 & 0.99894 & 0.99902 & 0.66624 \\
\hline 
$y=7/2$ & 0.66567 & 0.99806 & 0.99902 & 0.99806 & 0.66567 \\
\hline
$y=9/2$ & 0.45985 & 0.66567 & 0.66624 & 0.66567 & 0.45985 \\
\hline
\end{tabular}
\caption{\it The values of $D_L(x)$ for $\delta p/\nu\rho$ set to 1}
\label{Tdel}
\end{table}

\begin{figure}[hbt]
\label{fdel}
\centering
\epsfig{file=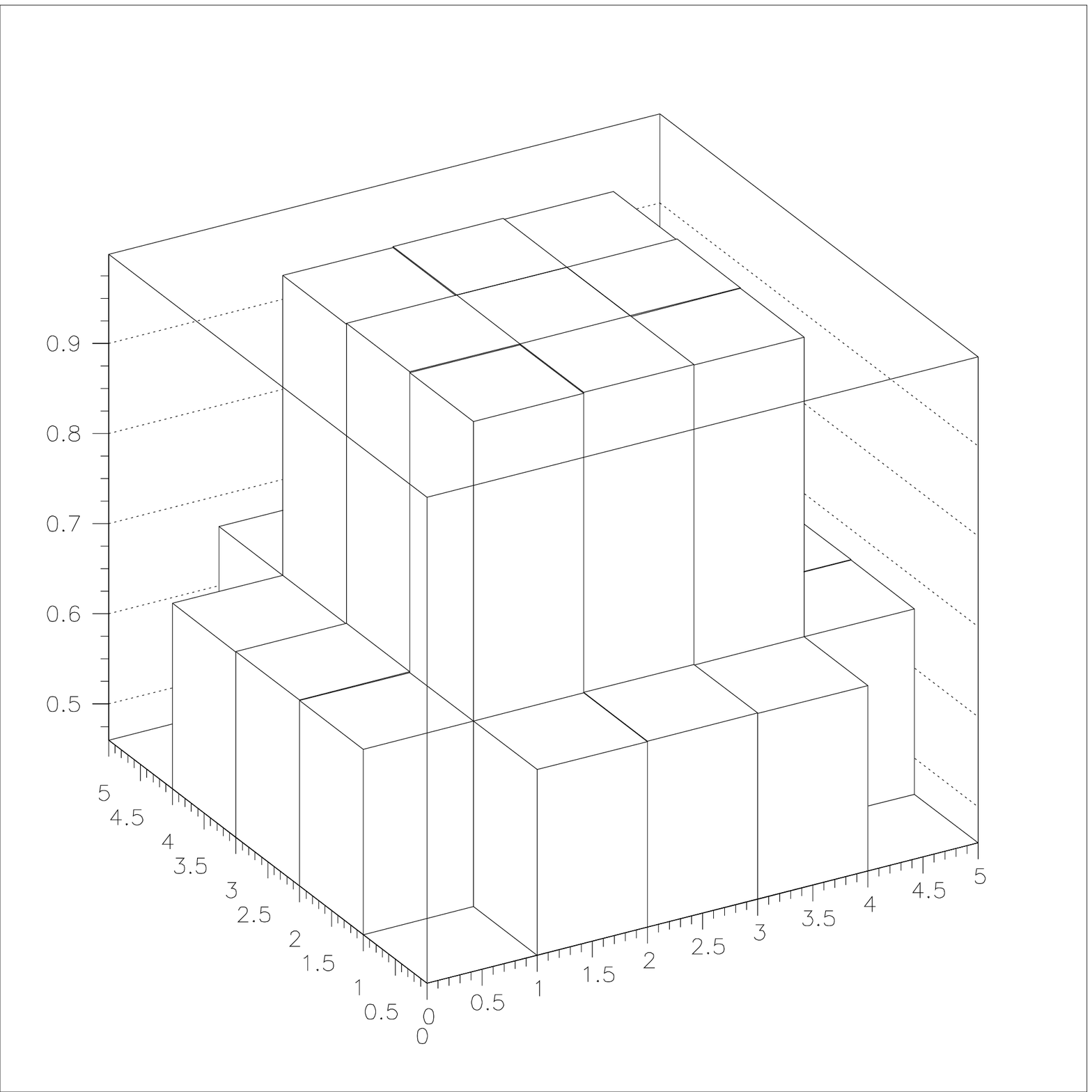,width=420pt}
\caption{\it Values of $D_L(x)$}
\end{figure}

\section{Invariance of the lattice equations of motion under a factor 2 RGT}

We would like to show explicitly that the lattice equations of motions
remain unchanged under a factor 2 renormalization group transformation.  To
do this we must first find the recursion relation, relating the
equations on the coarse lattice to the ones on the fine lattice. 
To find the recursion relation, we add to the action 
in the fine lattice variables $V_{d,x}$ (with spacing $a$) the constraint 
involving the coarse lattice variables $V'_{d,x'}$ (with spacing $2a$),
\begin{equation}
V'_{d,x'} = \frac{1}{2^{d-1}} \sum_{x \in x'} V_{d,x},
\end{equation}
using a Lagrange multiplier $\eta_{d,x'}$.  
Varying the action with respect to the various fields yields equations 
analogous to the ones relating the continuum fields to the lattice
fields
\begin{eqnarray}
\label{recursion}
\sum_y V_y \omega_{dd}(y,x) &+& \frac{\delta p}{ \nu \rho}D_L(x) -
\frac{1}{2^{d-1}} \sum_{x} \theta_{d,x'} \eta_{d,x'}(x) = 0 ,\nonumber \\
V'_{d,x'} &-&\frac{1}{2^{d-1}} \sum_{x \in x'} V_{d,x} - 
\alpha_2 \eta_{d,x'}=0 , \nonumber \\
\eta_{d,x'}&=&0.
\end{eqnarray}
Here $\theta_{d,x'}(x) = 1$ for $x \in x'$ and zero otherwise and
$\alpha_2$ is a constant to be adjusted to fit with our earlier choice of
optimization parameter $\alpha$.  Fourier 
transforming the above equations gives
\begin{eqnarray}
&&V_d(k) \omega_{dd}(k) + \frac{\delta p}{ \nu \rho} \omega_{dd}(k) \Delta_L(k) -
\tilde \Pi(k) \eta_d(k) = 0 , \nonumber \\
&&V'_d(k') - \sum_{m \in \{0,1\}^{d-1}} V_d(k' + \pi m/a) \tilde \Pi(k' + \pi m/a) 
\prod_{i=1}^{d-1} (-1)^l_i - \alpha_2 \eta(k') =0 ,\nonumber \\
\eta(k')&=&0 ,
\end{eqnarray}
where 
$k'_i = \pi/L n_i$
with $n_i \in \{1,N/2\}$ only, and
\begin{equation}
\tilde \Pi(k) = \prod_{i=1}^{d-1} \cos(k_i a/2) =  \frac{\Pi(2k)}{\Pi(k)}
\end{equation}
is the Fourier transform of $\theta_x'(x)$.  We note that in the above
equations $\eta_d(k')$ naturally extends to momenta $n_i>N/2$ by 
\begin{equation}
\eta_d(k'+ \pi m_i \hat i/a) = (-1)^{m_i} \eta_d(k').
\end{equation}
Solving the first equation for $V_d(k)$ we obtain
\begin{equation}
V_d(k) = \frac{1}{2^{d-1}} \omega(k)^{-1} \tilde \Pi(k) 
- \frac{\delta p}{ \nu \rho} \Delta_L(k).
\end{equation}

Inserting this relation into the other two equations and
comparing with eq.(\ref{perfecteq}) yields recursion formulas
for $\omega_{dd}(k)$ and $\Delta_L(k)$:
\begin{eqnarray}
\omega'_{dd}(k')^{-1}&=& \sum_{m \in \{0,1\}^{d-1}}
\omega_{dd}(k' + \pi m/a)^{-1} \tilde \Pi(k'+ \pi m/a)^2 + \alpha_2 \nonumber \\
\Delta'_L(k')&=& \sum_{m \in \{0,1\}^{d-1}} \Delta_L(k' + \pi m/a) 
\tilde \Pi(k' + \pi m/a) (-1)^{\sum_i m_i}. 
\end{eqnarray}
We now plug in the explicit form of $\omega_{dd}(k)$ from eq.(\ref{omega}) into
the first recursion relation
\begin{eqnarray}
\omega'_{dd}(k')^{-1}&=& \sum_{m \in \{0,1\}^{d-1}}  
\sum_{l \in \Z^{d-1}}\frac{1}{(k' + \pi (2l + m)/a)^2}
\Pi(2k' + 2 \pi (2l + m)/a)^2 \nonumber \\
&+&a^2 \alpha \frac{\Pi(2(k'+\pi m))^2}{\Pi(k'+\pi m)^2} 
+ \alpha_2 \nonumber \\
&=&\sum_{l' \in \Z^{d-1}}\frac{1}{(k' + 2 \pi l'/a')^2} \prod_{i=1}^{d-1} 
\frac{4 \sin^2(a'k'_i + 2 \pi l'_i)}{a'k'_i + 2 \pi l'_i} +
a^2 \alpha + \alpha_2 ,
\end{eqnarray} 
where we have let $l'_i = 2l_i + m_i$ and $a' = 2a$ in the last step.
Now, letting $\alpha_2=3a^2\alpha$ we recover the familiar form for
$\omega'_{dd}(k')$. 
Thus, we see that under the factor 2 recursion relation, 
$\omega_{dd}(k)$ for lattice
spacing $a$ transforms into $\omega'_{dd}(k')$ for lattice spacing $2a$.  It
can be shown that $\Delta_L(k)$ behaves similarly.  The fact that the
equations transform properly indeed confirms the they are
perfect lattice equations of motion.  

The recursion relation also serves as a powerful tool for numerical
methods because of the locality of the couplings.  For example, one
cannot solve analytically the lattice equations for a rectangular cross
section from which a smaller rectangular slice has been removed.
However, eq.(\ref{recursion}) is not specific to a given channel
geometry and hence it still applies.  To find the coupings for inward corner
points one need only choose a very fine lattice.  On it specify the couplings
already known, make ultralocal guesses for the unknown sites, and
perform several recursions.  Since the couplings are so local, if the
lattice is fine enough they serve as a very good approximation to the
actual values that define the equations of motion.  Thus, the averaging
process of the recursive step will tend to converge to fixed values,
thereby providing the necessary information for the unknown sites.  In
fact, we could have derived all the couplings in the above tables in
this fashion by making ultralocal naive guesses for the Poissonian and
then performing recursions.

In addition, we have numerically calculated the averaged velocity field
for $N=6$, $a=1$ and for $N=3$, $a=2$. This
was done by first numerically transforming $\omega_{dd}(k)$ and
$\omega_{dd}(k)\Delta_L(k)$ back into coordinate space. The values obtained
were then used to solve eq.(\ref{coordseqn}) for the lattice velocity
field.  The solution was found iteratively by making an initial
guess for the field values and then using the equation of motion to find
the field at $x$ in terms of the fields at $x' \neq
x$. With each iterative step the velocity field at $x$ was replaced (if
different) by
the value of the field as found through the equation of motion.  The 
iteration was stopped once the field converged to
stationary values.  For the above cases the field values are provided in
tables \ref{3T}, \ref{6T} and plotted in figures 6 and 7. We have explicitly checked that a factor 2 RGT on the fine
solution gives the coarse lattice fields values.  The fact that they satisfy 
eq.(\ref{constraint}) confirms that the equations of motion found 
are perfect.
\begin{table}
\centering
\begin{tabular}{|c|c|c|c|}
\hline
* & $x=1/2$ & $x=3/2$ & $x=5/2$ \\
\hline
$y=1/2$ & 0.81972 & 1.40674 & 0.81972  \\
\hline
$y=3/2$ & 1.40674 & 2.48867 & 1.40674  \\
\hline
$y=5/2$ & 0.81972 & 1.40674 & 0.81972  \\
\hline 
\end{tabular}
\caption{\it Velocity field values for $N=3$}
\label{3T}
\end{table}
  
\begin{figure}[hbt]
\label{3flow}
\centering
\epsfig{file=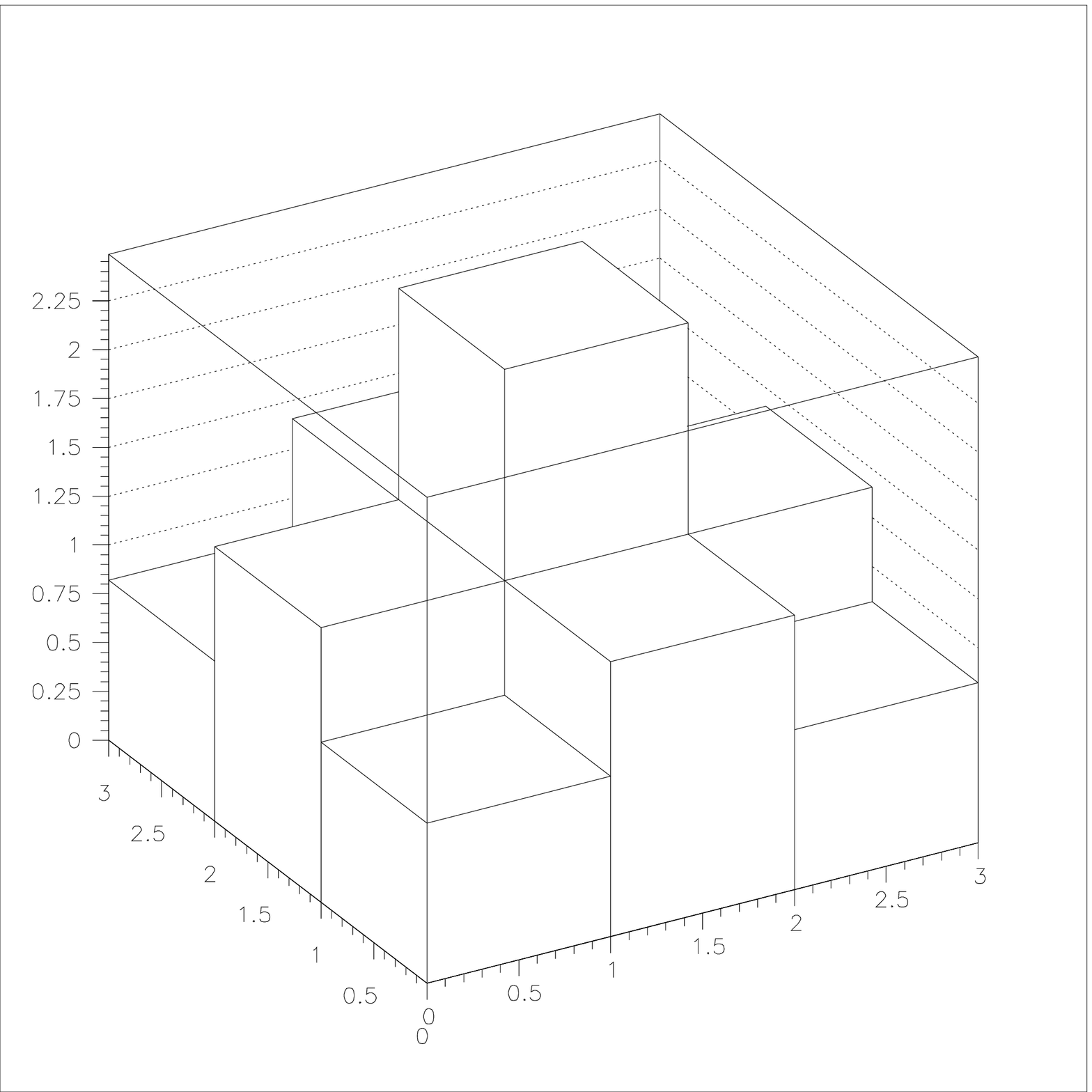, width=420pt}
\caption{\it The velocity field for $N=3$}
\end{figure}

\begin{table}
\centering
\begin{tabular}{|c|c|c|c|c|c|c|}
\hline
* & $x=1/2$ & $x=3/2$ & $x=5/2$ & $x=7/2$ & $x=9/2$ & $x=11/2$ \\
\hline
$y=1/2$ & 0.31513 & 0.68669 & 0.83510 & 0.83510 & 0.68669 & 0.31513 \\
\hline
$y=3/2$ & 0.68669 & 1.58922 & 1.97725 & 1.97725 & 1.58992 & 0.68669 \\
\hline
$y=5/2$ & 0.83510 & 1.97725 & 2.48758 & 2.48758 & 1.97725 & 0.83510 \\
\hline 
$y=7/2$ & 0.83510 & 1.97725 & 2.48758 & 2.48758 & 1.97725 & 0.83510 \\
\hline
$y=9/2$ & 0.68669 & 1.58922 & 1.97725 & 1.97725 & 1.58992 & 0.68669 \\
\hline
$y=11/2$ & 0.31513 & 0.68669 & 0.83510 & 0.83510 & 0.68669 & 0.31513 \\
\hline
\end{tabular}
\caption{\it Velocity field for $N=6$}
\label{6T}
\end{table}

\begin{figure}[hbt]
\label{6flow}
\centering
\epsfig{file=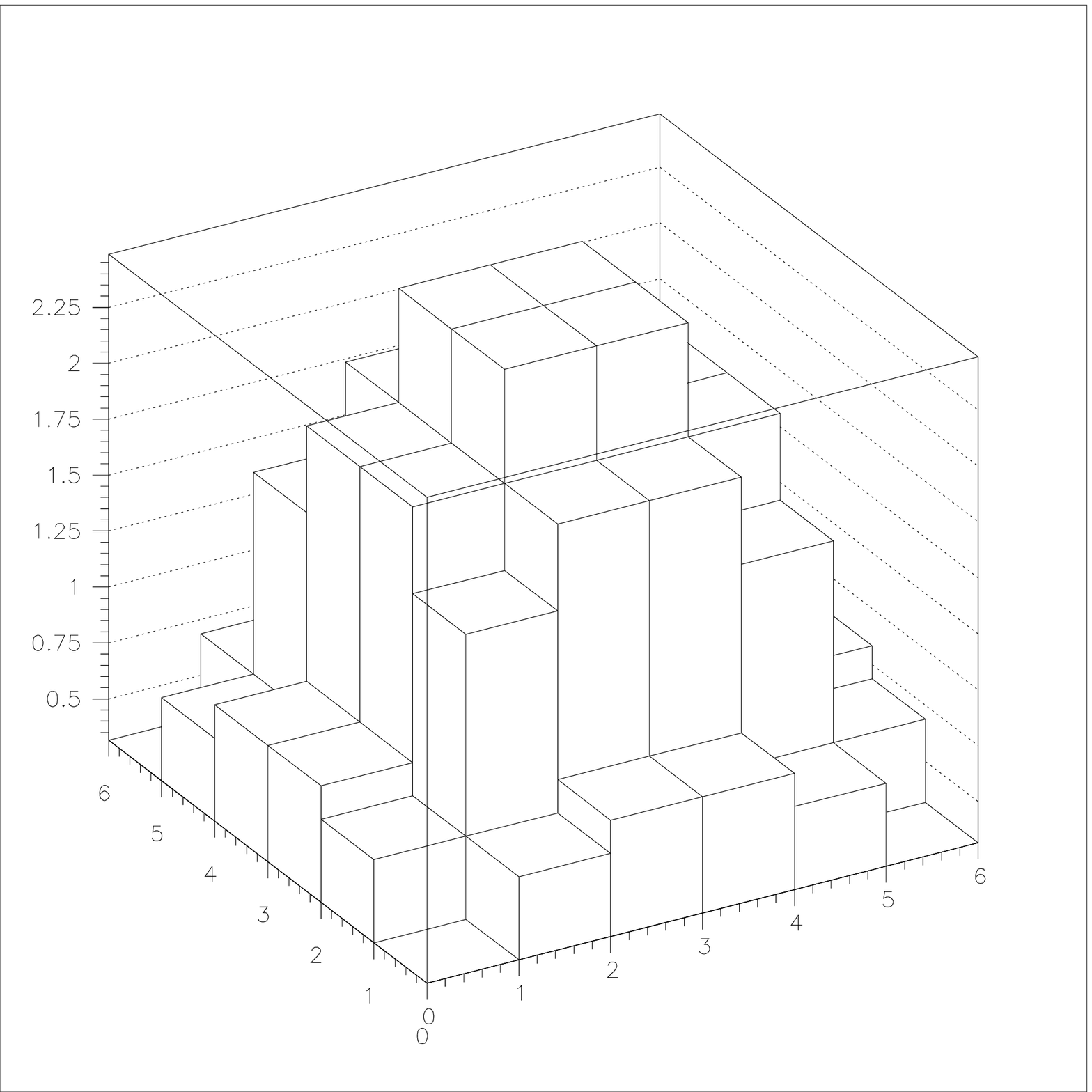, width=420pt}
\caption{\it The velocity field for $N=6$}
\end{figure}

\section{The perfect velocity field}

Up to now we have shown that the perfect lattice equations of motion give
exact results for the continuum fields averaged over cubes. In many practical
cases one is most interested in these averaged quantities. Still, in some cases
one would also like to know the values of the continuum fields themselves. In
particular, when one works on a very coarse lattice the averaged lattice
quantities may not contain enough information, and one would like to extract
information for continuum points.
Fortunately, this is possible with a perfect discretization. In fact, one can
reconstruct the continuum field from the lattice data. For this purpose we
insert eq.(\ref{etaequation}) into eq.(\ref{vequation}) and obtain
\begin{equation}
v_d(k) = \frac{1}{k^2}[\Pi_d(k) \omega_{dd}(k)(\frac{1}{a^2} V_d(k) +
\frac{\delta p}{\nu \rho} \Delta_L(k)) - \frac{\delta p}{\nu \rho} \delta_L(k)],
\end{equation}
which gives the continuum velocity field in terms of the lattice field $V_d$.
Transforming back to coordinate space one can reconstruct exactly the
continuum velocity field. When one considers the continuum velocity field at
lattice points one finds
\begin{equation}
v_d(x) = \sum_{x'} \rho_{dd}(x,x') V_{d,x'} + 
\frac{\delta p}{\nu\rho} \lambda_d(x).
\end{equation}
For a 2-d system $\rho_{dd}$ and $\lambda_d$ are
ultralocal and
\begin{equation}
v_d(x) = \frac{1}{8}(V_{d,x+a} + 6 V_{d,x} + V_{d,x-a}) -
\frac{1}{3} \frac{\delta p}{\nu\rho}
\end{equation}
for $x \neq a/2,(N-1/2)a$ and for the boundary point $x = a/2$
\begin{equation}
v_d(a/2) = \frac{1}{8}(5 V_{d,a/2} + V_{d,3a/2}) - 
\frac{1}{3} \frac{\delta p}{\nu\rho}.
\end{equation}
For continuum points which are not lattice points one can also obtain the
values of $\rho_{dd}$ and $\lambda_d$ by Fourier transformation.

\section{Perfect discretization for static slow flows}

Now we turn to general static flows with not necessarily constant pressure
gradients. In this case the Navier-Stokes equation reduces to
\begin{equation}
\partial_j \partial_j v_i(y) = \frac{1}{\nu\rho} \partial_i p(y).
\end{equation}
For reasons of simplicity we restrict ourselves to an infinite volume. The
inclusion of boundary conditions is straightforward (though tedious) and can
be done in analogy to the Hagen-Poiseuille flow. Although the above equation
still follows from an action principle, here we restrict ourselves to
working with the equations of motion alone. Eq.(\ref{Navierv}) then takes the
form
\begin{equation}
\partial_i \partial_i v_j(y) - \frac{1}{\nu\rho} \partial_j p +
\sum_x \eta_{j,x} \theta_{j,x}(y) = 0,
\end{equation}
and we still have
\begin{equation}
V_{j,x} - \frac{1}{a^{d-1}} \int_{f_{j,x}} d^{d-1}y \ v_j(y) - 
a^2 \alpha \eta_{j,x} = 0, \ \eta_{j,x} = 0.
\end{equation}
Similarly, the continuity equation leads to
\begin{equation}
\frac{1}{\nu\rho}\partial_i v_i(y) + \sum_x \eta_x \theta_x(y) = 0,
\end{equation}
and we also have
\begin{equation}
P_x - \frac{1}{a^d} \int_{c_x} d^dy \ p(y) - a^2 \beta \eta_x= 0, \ \eta_x = 0.
\end{equation}
Note that we have included a parameter $\beta$ analogous to $\alpha$
to be optimized for locality later.
It will turn out that this is not necessary, because the equations of motion
can be optimized for locality without it.  Since we are now in an
infinite volume 
\begin{eqnarray}
v_j(k)&=&\int d^dy \ v_j(y) \prod_{i=1}^d \exp(-i k_i y_i),\\
p(k)&=&\int d^dy \ p(y) \prod_{i=1}^d \exp(-i k_i y_i),
\end{eqnarray}
where $k \in \R^d$.  Hence, it is convenient to now let 
\begin{eqnarray}
V_j(k)&=&a^{d-1} \sum_{x} V_{j,x} \prod_{i \neq j} \exp(-i k_i x_i),\\
P(k)&=&a^d \sum_{x} P_x \prod_{i=1}^d \exp(-i k_i x_i),
\end{eqnarray}
where $k \in [-\pi/a ,\pi/a]^d$.  We note that given the geometry of Fig.
1, if $P_x$ lie at points $x$ whose coordinates are integers, 
then $V_{j,x}$ will lie at a point $x$ whose $j$-th coordinate is a half
integer.
Also, as before, it is possible to extend the lattice fields to momenta outside
$[-\pi/a, \pi/a]$ by letting
\begin{eqnarray}
V_j(k+2 \pi l_i \hat i/a) &=& V_j(k) , i \neq j \nonumber \\ 
V_j(k+2 \pi l_j \hat j/a) &=& (-1)^{l_j} V_j(k),  \nonumber \\
P(k+2 \pi l_i \hat i/a) &=& P(k), \mbox{for all $i$}.
\end{eqnarray}
We are thus ready to go to Fourier space where the above equations take
the form
\begin{eqnarray}
\label{transnavier}
&&-k^2 v_j(k) - \frac{i}{\nu \rho} k_j p(k) +\Pi_j(k) \eta_j(k)=0 \\
\label{aveV}
&&V_j(k) - \sum_{l \in \Z^d} v_j(k+2 \pi l/a) \Pi_j(k+2 \pi l/a)(-1)^{l_j}
-a^2 \alpha \eta_j=0 \\
&&\frac{i}{\nu \rho} k_j v_j(k) + \Pi(k) \eta(k)=0 \\
\label{aveP}
&&P(k) - \sum_{l \in \Z^d} p(k+2 \pi l/a) \Pi(k+2 \pi l/a) -a^2 \beta \eta(k)=0.
\end{eqnarray}
Solving  for the continuum fields we obtain
\begin{eqnarray}
&&p(k) = -(\nu \rho)^2 \Pi(k) \eta(k) - \frac{i \nu \rho}{k^2} k_j
\Pi_j(k) \eta_j(k) \\
&&v_i(k) = \frac{i \nu \rho}{k^2} k_i \Pi(k) \eta(k) - \frac{k_i k_j
-k^2 \delta_{ij}}{k^4} \Pi_j(k) \eta_j(k).
\end{eqnarray}
Reinserting these back into eq.(\ref{aveV}) and eq.(\ref{aveP}) gives 
equations of motion for the
lattice fields and the Lagrange multiplier fields
\begin{eqnarray}
&&V_i(k) - i\nu \rho \omega_i(k) \eta(k) +\omega_{ij}(k)
\eta_{j}(k)=0 \nonumber \\
&&P(k) + (\nu \rho)^2 \omega(k) \eta(k)+ i \nu \rho \omega_j(k)
\eta_j(k)=0.
\end{eqnarray}
Here we have introduced the functions
\begin{eqnarray}
&& \omega(k)=\sum_{l \in \Z^d} \Pi(k+2 \pi l/a)^2 + \beta = 1 + \beta \\
&& \omega_i(k)=\sum_{l \in \Z^d} \frac{k_i +2 \pi l_i/a}{(k+2 \pi
l/a)^2} \Pi_i(k + 2 \pi l/a) \Pi(k + 2 \pi l/a)(-1)^{l_i} \\
&& \omega_{ij}(k)^{-1}=\sum_{l \in \Z^d} \frac{(k_i +2 \pi l_i/a)(k_j +
2 \pi l_j/a)-(k+2 \pi l/a)^2 \delta_{ij}}{(k+2 \pi
l/a)^4} \nonumber \\ 
&& \times \Pi_i(k + 2 \pi l/a)\Pi_j(k + 2 \pi l/a)(-1)^{l_i+l_j}+ a^2 \alpha .
\end{eqnarray}
We can rewrite the above equations in a simple fashion by letting
$V_0=P$ and $\eta_0=\eta$
\begin{equation}
V_{\mu}=\Omega_{\mu \nu}^{-1} \eta_{\nu},
\end{equation}
where $\Omega_{\mu \nu}$ is defined through the above relations.
Finally, we solve for the Lagrange multiplier fields and set them equal
to zero to obtain the equations of motion for the lattice fields
\begin{equation}
\label{eqofmotionnavier}
\Omega_{ \mu \nu} V_{\nu}=0.
\end{equation}
We must now optimize for locality.  As before, we go to $d=1$ and choose
$\alpha$ and $\beta$ so that the equations of motion take an ultralocal
form.  In one dimension eq.(\ref{eqofmotionnavier}) becomes
\begin{equation}
\frac{1}{(1+\beta)\alpha-1/(2\sin(k/2))^2}
\left( \begin{array}{cc}
\alpha & i/(2\sin(k/2)) \\
-i/(2\sin(k/2) &  
1+\beta
\end{array} \right) 
\left( \begin{array}{c} P \\ V \end{array} \right) = 0. 
\end{equation}
Setting $\alpha=0$ and $\beta=0$, we obtain familiar ultralocal
equations
\begin{eqnarray}
&& 2\sin(k/2) V(k)=0 \\
&& 2 i \sin(k/2) P(k) = (2\sin(k/2))^2 V(k).
\end{eqnarray}
Having solved this simplified version of the Navier-Stokes equations we 
proceed to add the nonlinear term.

\section{Perturbative treatment of the nonlinear term}
Since the nonlinear term of the Navier-Stokes equation is quadratic in the
velocity, it is possible to solve the lattice equations of motion only
perturbatively.  Here we demonstrate how to do this for the first order
correction.  We begin by introducing the modified version of
eq.(\ref{transnavier}), now with the new term  
\begin{equation}
-\frac{i}{\nu} v_j(k) k_j v_i(k) -k^2 v_i(k) -\frac{i}{\nu \rho} k_i v_i(k) +
\Pi_j(k) \eta_j(k) = 0.
\end{equation}
Because we are only interested in the first order correction we need only
keep terms in the nonlinear contribution that are quadratic in the
Lagrange multiplier
fields. Thus, for the nonlinear term, it is
sufficient to express the continuum fields as linear functions of the
Lagrange fields.  We make the following ansatz for $p^{(2)}$
and $v_i^{(2)}$, the second order contributions:
\begin{eqnarray}
&&p^{(2)} = A \eta(k)^2 + \eta(k) B_j(k) \eta_j(k) + \eta_i(k) C_{ij}(k)
\eta_j(k) \\
&&v_i^{(2)} = D_i(k) \eta(k)^2 + \eta(k) E_{ij} \eta_j(k) + F_{ijk}
\eta_j(k) \eta_k(k).
\end{eqnarray}
Then $p^{(2)}$ and $v_i^{(2)}$ must obey the following equations,
\begin{eqnarray}
&&\frac{i \nu \rho^2}{k^2}k_i \Pi(k)^2 \eta^2-\eta 
\frac{\rho (k_i k_j-k^2\delta_{ij})}{k^4}\Pi(k)\Pi_j(k) \eta_j -k^2
v_i^{(2)}-\frac{i}{\nu\rho} k_i p^{(2)}=0. \\
&& k_i v_i^{(2)}(k) = 0.
\end{eqnarray}
Solving them, we obtain
\begin{eqnarray}
&&p^{(2)}(k) = \nu^2\rho^3 \frac{\Pi(k)^2}{k^2} \eta(k)^2 \\
&&v_i^{(2)}(k) = \rho \eta(k) \frac{k_i k_j - k^2 \delta_{ij}}{k^6} \Pi_j(k)\Pi(k) \eta_j(k).
\end{eqnarray}
This procedure can now be repeated for the lattice fields.  We make the
ansatz,
\begin{equation}
\eta_{\lambda}^{(2)}(k)=V_{\mu}(k) A_{\mu\nu}^{\lambda}(k) V_{\nu}(k),
\end{equation}
where again $V_0=P$.  Keeping only the second order terms, yields
\begin{eqnarray}
&& -i\nu\rho\omega_i V_{\mu} A_{\mu\nu}^{0} V_{\nu}+\omega_{ij}(k)
V_{\mu} A_{\mu\nu}^{j} V_{\nu} 
-\rho V_{\mu} \Omega_{\mu 0}^{T} S_{ij} \Omega_{j\nu} V_{\nu} = 0 \\
&& (\nu\rho)^2 \omega V_{\mu} A_{\mu\nu}^{0} V_{\nu}+i\nu\rho\omega_j V_{\mu} A_{\mu\nu}^{j} V_{\nu}
-\nu^2\rho^3 V_{\mu} \Omega_{\mu 0}^{T} S \Omega_{0 \nu} V_{\nu} = 0.
\end{eqnarray}
Here, the following functions have been introduced
\begin{eqnarray}
&& S(k)=\sum_{l \in \Z^d} \frac{1}{(k+2 \pi
l/a)^2} \Pi_i(k + 2 \pi l/a)^3  \\
&& S_{ij}(k)=\sum_{l \in \Z^d} \frac{(k_i +2 \pi l_i/a)(k_j +
2 \pi l_j/a)-(k+2 \pi l/a)^2 \delta_{ij}}{(k+2 \pi
l/a)^6} \nonumber \\ 
&& \times \Pi_i(k + 2 \pi l/a)\Pi_j(k + 2 \pi l/a)\Pi(k + 2 \pi l/a)(-1)^{l_i+l_j}.
\end{eqnarray}
We rewrite the above equations in a more compact form
\begin{equation}
\Omega^{-1}_{\sigma\lambda} A_{\mu\nu}^{\lambda}=M_{\mu\nu}^{\sigma}.
\end{equation}
This, finally gives the first order nonlinear equation of motion
\begin{equation}
\Omega_{\mu\nu} V_{\nu} + \Omega_{\mu\nu} V_{\lambda}
M_{\lambda\sigma}^{\nu} V_{\sigma} = 0.
\end{equation}
In principal this procedure could be repeated to obtain non-linear lattice
equations up to any order in the lattice fields.  

\section{Conclusions}

In this paper we have performed the first steps towards a perfect
discretization of the Navier Stokes equations. The essential idea is to
define coarse grained variables by averaging the continuum pressure and
velocity fields over appropriate spatial regions. The corresponding averaged
field variables naturally live on a lattice. In the cases of Hagen-Poisseille
flow and slow static flows in general we derived the exact equations of motion
for the lattice variables, and hence a perfect discretization of the 
corresponding continuum equations. The incorporation of boundary conditions is
nontrivial, and has been done explicitly for Hagen-Poisseille flow. For
practical applications it is essential that the perfectly discretized equations
of motion are very local (although in general not ultralocal). In particular,
the lattice couplings to distant variables decay exponentially, and one can
safely restrict oneself to a few nearby neighbors. In fact, we have performed
numerical simulations using the perfectly discretized equations of motion for
laminar flow in a channel with quadratic cross section and we have verified 
explicitly that an arbitrarily coarse lattice indeed gives continuum answers. 
Our method can be directly applied to more general geometries, although 
non-cubic lattices require nontrivial modifications.
An important aspect of our construction are the so-called perfect fields.
Indeed our method allows to interpolate the continuum fields from the lattice
data. Hence, especially when we are working on a coarse lattice, we are not
limited to lattice points.

Of course, a method like that would be most welcome in numerical simulations
of more complicated fluid dynamics systems, especially in the case of 
turbulence. In these simulations the effects of a finite lattice spacing are 
the main source of systematic errors. We have made a first step in this
direction by deriving the perfect equations of motion in the presence of
the nonlinear term that gives rise to turbulence. However, we have included
that term only to first order, while a perfect discretization of turbulent
flows would require a treatment to all orders. This could be achieved 
numerically in analogy to \cite{Has94} by solving a minimization problem on
a multigrid. When one wants to simulate turbulence one should not restrict
oneself to static flows. Then the definition of the averaged lattice variables
should be generalized. In particular, one should also average pressure and
velocity in time. In addition, one may want to simulate compressible fluids,
which would bring in an averaged density variable. Our techniques are still
applicable to these cases, but the actual implementation is nontrivial.

In conclusion, we have proposed a renormalization group treatment of problems
in fluid dynamics. The renormalization group implies that perfect 
discretizations of the continuum Navier-Stokes equations exist. The
perfectly discretized equations of motion have been constructed explicitly
for slow static flows, and have been optimized for locality. This is essential
in practical applications. The renormalization group offers a systematic way
of treating small scale effects in fluid dynamics. Realizing this program for
the full Navier-Stokes problem requires a lot of work. Still, the first steps
that we have taken, have led to promising results.

\section*{Acknowledgments}
We would like to thank W. Evers for very important discussions during
the initial stages of this project.  We would also like to thank
W. Bietenholz for very useful comments and remarks about the paper.

\end{document}